\documentclass[usenames,dvipsnames,a4paper,11pt]{article}
\usepackage[margin=2.5cm]{geometry}
\usepackage{graphicx}
\usepackage{epsfig}
\usepackage{amsmath}
\usepackage{amssymb}
\usepackage{array}
\usepackage{braket}
\usepackage{scalerel}
\usepackage{cite}
\usepackage{hyperref}
\usepackage{cleveref}
\usepackage{float}
\usepackage[font={footnotesize,it}]{caption}
\usepackage{authblk}
\usepackage[utf8]{inputenc}
\usepackage{cancel}
\usepackage{tikz}
\usepackage{color}
\usepackage{hyperref}
\hypersetup{
    colorlinks=true,
    linkcolor=blue,
    filecolor=red,      
    urlcolor=blue,
    citecolor=red
} 
\begin{document}
\begin{titlepage}

\title{Entanglement negativity at large central charge}

\author[1]{Vinay Malvimat\thanks{\noindent E-mail:~ vinaymm@iitk.ac.in }}
\author[2]{Gautam Sengupta\thanks{\noindent E-mail:~  sengupta@iitk.ac.in}}

\affil[1,2]{Department of Physics, Indian Institute of Technology Kanpur, Kanpur 208016, INDI{\cal A}}

\maketitle

\abstract{  

 We establish the large central charge behaviour of the entanglement negativity for a mixed state configuration of a single interval enclosed between two intervals in a holographic $CFT_{1+1}$. To this end we utilize the monodromy technique to elucidate the large central charge limit of a four point twist correlator which characterizes the negativity of the mixed state configuration. Subsequently we analyze the large central charge limit of the six point twist correlator which reduces to the above four point correlator in a specific limit.
The results provide a strong consistency check for our recently proposed holographic entanglement negativity conjecture in the $AdS_3/CFT_2$ scenario. 

}
\end{titlepage}
\tableofcontents
\newpage

\definecolor{Nblue}{rgb}{0.0, 0.53, 0.74}
\section{Introduction}

Over the last decade significant advances in the understanding of quantum entanglement 
in the framework of the AdS-CFT correspondence has revealed a remarkable connection with issues of quantum gravity and the structure of space time \cite{VanRaamsdonk:2009ar,PhysRevD.86.065007,VanRaamsdonk:2010pw}. In this context 
{\it entanglement entropy} has played a key role as the measure to characterize quantum entanglement for bipartite pure states. For systems with finite degrees of freedom this measure may be computed directly from its definition. However, for extended quantum systems the computation of entanglement entropy is arduous as it involves the evaluation of the eigenvalues of an infinite dimensional density matrix. For $(1+1)$-dimensional conformal field theories ( $CFT_{1+1}$ ) this issue was addressed through a {\it replica technique} which was developed in a series of interesting articles by Calabrese et. al\cite{Calabrese:2009qy,Calabrese:2004eu}. Subsequently in a significant communication, Ryu and Takayanagi advanced a holographic conjecture to compute entanglement entropy of bipartite states in conformal field theories through the corresponding dual bulk  geometry. According to this the entanglement entropy for a 
subsystem in a holographic $CFT_d$ was proportional to the area of the co dimension two bulk static minimal surface homologous to the subsystem, in the dual $AdS_{d+1}$ geometry\cite{Ryu:2006bv,Ryu:2006ef}. A covariant version of this conjecture for $CFT_d$s dual to non static bulk $AdS_{d+1}$ configurations was subsequently communicated in \cite{Hubeny:2007xt}. 

In the $AdS_3/CFT_2$ scenario Ryu and Takayangi \cite{Ryu:2006bv,Ryu:2006ef} showed that for a single interval in a $CFT_{1+1}$ the holographic entanglement entropy exactly reproduced the result of \cite{Calabrese:2009qy,Calabrese:2004eu} in the large central charge limit. The case for multiple disjoint intervals was however more complex as this 
was related to the higher point twist correlators in the $CFT_{1+1}$ and in general was dependent on the full operator content of the theory. In this connection it is useful to recollect that the semi classical bulk gravitational description in the limit $G_N\to0$ corresponds to the large central charge $c\to\infty$ limit of the dual 
$CFT_{1+1}$ in the $AdS_3/CFT_2$ scenario. The authors in the articles \cite{Headrick:2010zt,Hartman:2013mia} computed the entanglement entropy of two disjoint intervals through the relevant four point twist correlator employing the conformal block expansion. Remarkably they were able to show that the holographic entanglement entropy characterized by the Ryu Takayanagi conjecture exactly reproduced their results in the large central charge limit. 

The key technique used in \cite{Hartman:2013mia} derives from a result due to Zamolodchikov et al. \cite{Zamolodchikov:1995aa} which states that the conformal blocks of the four point correlator in the Liouville theory exponentiate in the large central charge limit. Subsequently Hartman in \cite{Hartman:2013mia}, employed the monodromy technique for the determination of the conformal blocks for the higher point correlators of the twist fields in the $CFT_{1+1}$. Notably this monodromy technique has been very useful recently to investigate general higher point correlators in $CFT_{1+1}$ related to a variety of significant issues involving thermalization and the information loss paradox \cite{Fitzpatrick:2014vua,Perlmutter:2015iya,Fitzpatrick:2015zha,Alkalaev:2015lca,Alkalaev:2016rjl,Fitzpatrick:2016ive,Anous:2016kss}. A proof of the
Ryu-Takayanagi conjecture in the context of the above $AdS_3/CFT_2$ scenario was developed in
\cite{Faulkner:2013yia}. The corresponding proof for the more general $AdS_{d+1}/CFT_d$ scenario was obtained in \cite {Lewkowycz:2013nqa} and the covariant version in \cite{Dong:2016hjy}.

In quantum information theory it is well known that entanglement entropy ceases to be a valid measure for the characterization of mixed state entanglement as it receives contributions from correlations which are irrelevant for the specific mixed state configuration\footnote{For example for the mixed state described by a finite temperature $CFT_{1+1}$ case it receives contributions from both classical and quantum correlations\cite{Calabrese:2009qy,Calabrese:2004eu,Calabrese:2014yza}.}. A significant resolution of the issue of mixed state entanglement in quantum information theory was advanced by Vidal and Werner in \cite{PhysRevA.65.032314}. In this article the authors introduced a computable measure 
termed {\it entanglement negativity } which characterized the upper bound on the 
{\it distillable entanglement} for a mixed state. The specification of the entanglement negativity involves the procedure of {\it purification} in which the bipartite system is embedded in a larger auxiliary system such that the composite is in a pure state and then tracing over the auxiliary system. Recently in a series of articles the authors in  \cite{Calabrese:2012ew,Calabrese:2012nk,Calabrese:2014yza} developed a replica technique for the entanglement negativity of bipartite pure and mixed states in $CFT_{1+1}$. In this context the large central charge limit for the entanglement negativity between two disjoint intervals in a $CFT_{1+1}$ was obtained by Kulaxizi et al. in\cite{Kulaxizi:2014nma}. In this article the authors employed the monodromy technique mentioned earlier to determine the conformal blocks for the relevant four point twist correlator. 

Recently, the present authors ( VM and GS) in a collaboration (CMS) advanced a holographic entanglement negativity conjecture for bipartite pure and mixed states in holographic CFTs\cite{Chaturvedi:2016rcn,Chaturvedi:2016rft,Chaturvedi:2016opa}. The holographic entanglement negativity was observed to be proportional to a specific algebraic sum of the areas of co dimension two bulk extremal surfaces (geodesic lengths in the $AdS_3$) anchored on the corresponding subsystems that reduces to a sum of the holographic mutual information between the subsystems. For the $AdS_3/CFT_2$ scenario this conjecture exactly reproduced the universal part of the corresponding $CFT_{1+1}$ results computed through the replica technique\footnote {The universal part of the entanglement negativity refers to the component which is independent of the detailed operator content of the CFT.}. One of the motivation for the present article is to demonstrate that the universal part of the $CFT_{1+1}$ result is the dominant contribution to the entanglement negativity in the large central charge limit. 
 
More recently, the present authors (VM and GS) in another collaboration proposed a distinct holographic entanglement negativity conjecture for mixed states of adjacent subsystems in CFTs \cite {Jain:2017aqk,Jain:2017xsu}. Interestingly this involved another specific algebraic sum of the areas of co dimension two bulk extremal surfaces anchored on the specific subsystems which reduced to the holographic mutual information between them. In this case for the $AdS_3/CFT_2$ scenario the holographic entanglement negativity following from our conjecture also exactly reproduced the corresponding $CFT_{1+1}$ results in the large central charge limit described in \cite{Kulaxizi:2014nma}.
 
Quite naturally these holographic conjectures and the above discussions makes it imperative to comprehensively investigate the large central charge limit of the entanglement negativity for mixed states in holographic $CFT_{1+1}$s. As stated above for the mixed state of adjacent intervals this issue was studied in detail by Kulaxizi et. al in \cite{Kulaxizi:2014nma}.
In the present article we address the corresponding problem for a distinct mixed state of a $CFT_{1+1}$ which  has crucial  relevance for the holographic conjecture described earlier \cite{Chaturvedi:2016rcn,Chaturvedi:2016rft,Chaturvedi:2016opa}. This mixed state configuration is described by a single interval ($A$) enclosed between two other intervals ($B_1\cup B_2$) in a $CFT_{1+1}$ as depicted in fig.(\ref{fig4}). A bipartite limit may then be taken to obtain the entanglement negativity corresponding to the full system $A\cup A^c$ as described in \cite{Chaturvedi:2016rcn,Chaturvedi:2016rft,Chaturvedi:2016opa}.

Motivated by the considerations described above, here we present a detailed investigation of this issue. In this context we utilize the monodromy technique mentioned earlier to examine the large central charge behavior of a four point twist correlator in a holographic $CFT_{1+1}$, which characterizes the entanglement negativity of the above configuration.We observe that the results obtained both in the $s$ and the $t$ channels are consistent with our holographic conjecture \cite {Chaturvedi:2016rcn} for the $AdS_3/CFT_2$ scenario. Subsequently we analyze the large central charge behaviour of a six point twist correlator which reduces to the above four point correlator in a specific limit. Through our analysis we are able to demonstrate that the universal part of the entanglement negativity dominates in the large central charge limit. 
Hence the results we obtain here, serve as a strong consistency check for our holographic  entanglement negativity conjecture for the above mentioned mixed state configuration,  as described in \cite {Chaturvedi:2016rcn,Chaturvedi:2016opa}. 

The article is organized as follows. In section two we briefly review the computation of the entanglement negativity of the relevant mixed state in a $CFT_{1+1}$. Subsequently in Section three we briefly review our holographic negativity conjecture in the context of the $AdS_3/CFT_2$ scenario. In the fourth section we present a concise account of the monodromy technique to obtain the conformal blocks for the four point correlator characterizing the negativity in a $CFT_{1+1}$ in the large central charge limit. In section five we utilize this technique to investigate the large-c behavior of the  four point twist correlator and demonstrate that the result in the relevant channels are consistent with our holographic conjecture. In Section six we analyze the large central charge behaviour of a six point twist correlator which reduces to the four point correlator in a specific limit. In the last section we present a summary of our results and  conclusions.

\section{Entanglement negativity in a $CFT_{1+1}$}\label{Rev}
In this section we briefly review the computation of the entanglement negativity for 
mixed states described by two different configurations in a $CFT_{1+1}$ relevant for our purpose. In this context we first introduce the definition of entanglement negativity in quantum information theory as proposed in \cite{PhysRevA.65.032314}.  The authors there considered a tripartite system in a pure state consisting of subsystems that are denoted as $A_1,A_2$ (such that $A_1\cup A_2=A$) and $A^c$ describing the rest of the system. The entanglement negativity characterizing the upper bound on the {\it distillable entanglement} between the subsystems $A_1$ and $A_2$ is defined as follows
\begin{equation}\label{endef}
 \mathcal{E} =  \log \mathrm{Tr}|(\rho_A^{T_2})|
\end{equation}
where the $\rho_A$ is the reduced density matrix of the subsystem $A=A_1\cup A_2$ and the superscript $T_2$ indicates the operation of partial transpose which is defined as follows 
\begin{equation}\label{part}
\langle e^{(1)}_ie^{(2)}_j|\rho^{T_2}|e^{(1)}_ke^{(2)}_l\rangle = 
\langle e^{(1)}_ie^{(2)}_l|\rho|e^{(1)}_ke^{(2)}_j\rangle .
\end{equation}
Here $|e^{(1)}_i\rangle$ and $|e^{(2)}_j\rangle$ represent the basis states of the subsystems $A_1$ and $A_2$ respectively.

\begin{figure}[H]
  \begin{center}
  \begin{tikzpicture}
  \draw[thick](-5,0)--(-3,0);
  \draw[red,thick] (-3,0)--(0.0,0); 
  \draw[thick] (0,0)--(1.8,0); 
  \draw[Nblue,thick] (1.8,0)--(4.8,0); 
  \draw[thick](4.8,0)--(6.5,0);
  \draw(-3,0.3)--(-3,-0.3);
  \draw(0.0,0.3)--(0.0,-0.3);
  \draw(1.8,0.3)--(1.8,-0.3);
  \draw(4.8,0.3)--(4.8,-0.3);
  \node()at (-1.6,0.5){$A_1$};
  \node()at (-3,-0.5){$u_1$};
  \node()at (3.4,0.5){$A_2$};
  \node()at (0.0,-0.5){$v_1$};
  \node()at (1.8,-0.5){$u_2$};
  \node()at (4.8,-0.5){$v_2$};
  \end{tikzpicture}
  \caption{Schematic of the configuration of the mixed state of two disjoint intervals $A_1$ and $A_2$}
  \label{fig1}
  \end{center}
\end{figure}
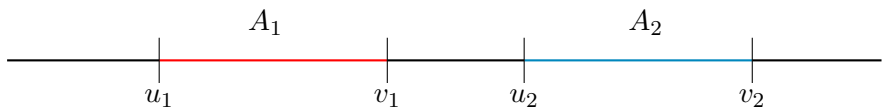

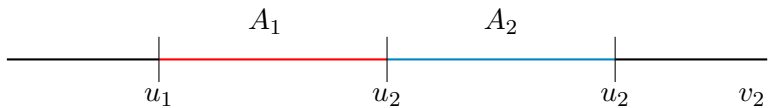
\begin{figure}[H]
  \begin{center}
  \begin{tikzpicture}
  \draw[thick](-5,0)--(-3,0);
  \draw[red,thick] (-3,0)--(0.0,0); 
  \draw[Nblue,thick] (0,0)--(3,0); 
  \draw[thick](3,0)--(5,0); 
  \draw(-3,0.3)--(-3,-0.3);
  \draw(0.0,0.3)--(0.0,-0.3);
  \draw(3,0.3)--(3,-0.3);
  \node()at (-1.6,0.5){$A_1$};
  \node()at (-3,-0.5){$u_1$};
  \node()at (1.5,0.5){$A_2$};
  \node()at (0.0,-0.5){$u_2$};
  \node()at (3,-0.5){$u_2$};
  \node()at (4.8,-0.5){$v_2$};
  \end{tikzpicture}
  \caption{Schematic of the configuration in the limit the two intervals $A_1$ and $A_2$ become adjacent}
  \label{fig2}
  \end{center}
\end{figure}

As discussed in the introduction, Calabrese et al. in \cite{Calabrese:2012ew,Calabrese:2012nk,Calabrese:2014yza} developed a replica technique to compute the entanglement negativity
for various pure and mixed state configurations in a $CFT_{1+1}$.  The first configuration depicted in fig.(\ref{fig1}) involves the subsystems $A_1$ and $A_2$ corresponding to two disjoint finite intervals denoted as$[u_1,v_1]$ and  $[u_2,v_2]$ respectively and $A^c$ describes the rest of the system. The replica definition of the entanglement negativity  between the subsystems $A_1$ and $A_2$ is given as follows
\begin{equation}\label{negrep}
\mathcal{E} = \lim_{n_e \rightarrow 1 } \log \mathrm{Tr}(\rho_A^{T_2})^{n_e}.
\end{equation}
Here $n_e$ denotes that the parity of the replica index $n$ is even. Note that this is because the definition of entanglement negativity given in eq.(\ref{endef}) matches with the above definition only if we assume the even parity of $n$ that is $n=n_e$ and then finally take the limit $n_e\to1$. Therefore the authors proposed the replica definition for the entanglement negativity as an analytic continuation of even sequences of $n$ to $n_e=1$ \footnote{Note that this analytic continuation is non-trivial and has been established only for expicit examples involving some simple conformal field theories \cite{Calabrese:2012ew,Calabrese:2012nk,Calabrese:2014yza}.}. The authors demonstrated that the quantity $\mathrm{Tr}(\rho_A^{T_2})^{n_e}$ in eq.(\ref{negrep}) is given by the following four point twist correlator 
\begin{equation}\label{4pt2int}
\mathrm{Tr}(\rho_A^{T_2})^{n_e} = 
\langle\mathcal{T}_{n_e}(u_1)\overline{\mathcal{T}}_{n_e}(v_1)\overline{\mathcal{T}}_{n_e}(u_2)\mathcal{T}_{n_e}(v_2)\rangle_{\mathbb{C}}.
\end{equation}
where ${\mathcal{T}}$ and $\overline{\mathcal{T}}$ are the twist and the anti-twist operators both of which have the scaling dimensions $\Delta_{n_e}=\frac{c}{24}(n_e-\frac{1}{n_e})$.
Note that the four point correlator in a $CFT_{1+1}$ can only be fixed upto a function of the cross ratio $\big[ x=\frac{(u_1-v_1)(u_2-v_2)}{(u_1-u_2)(v_1-v_2)}\big]$, which depends on the full operator content of the theory. This made the computation of the negativity in the case of the disjoint interval extremely complex except for some specific examples such as the free compactified boson and critical Ising model\cite{Calabrese:2009ez,Calabrese:2010he,Calabrese:2012nk}. However, in the limit of adjacent intervals described as $v_1\to u_2$ as depicted in fig.(\ref{fig2}), the four point twist correlator in eq.(\ref{4pt2int}) reduces to the following three point  twist correlator as follows
\begin{equation}\label{3ptadj}
\mathrm{Tr}(\rho_A^{T_2})^{n_e} = 
\langle\mathcal{T}_{n_e}(u_1)\overline{{\cal T}}^2_{n_e}(u_2)\mathcal{T}_{n_e}(v_2)\rangle_{\mathbb{C}}.
\end{equation}
where $\overline{{\cal T}}^2_{n_e}(u_2)$ corresponds to the twist operator which connects $j^{th}$-sheet of the Riemann surface to $(j-2)^{th}$-sheet and has the following scaling dimension 
\begin{equation}
\Delta_{n_e}^{(2)}=2\Delta_{\frac{n_e}{2}}=\frac{c}{12}(\frac{n_e}{2}-\frac{2}{n_e}.)
\end{equation}

The form of the three point function may be completely determined by the conformal symmetry up to a numerical constant ( which also depends on the full operator content) and leads to the following expression for the entanglement negativity \cite{Calabrese:2012ew,Calabrese:2012nk} of the mixed state in the $CFT_{1+1}$ at zero temperature.
\begin{equation}\label{enadj1}
{\cal E}=\frac{c}{4}\log[\frac{l_1l_2}{(l_1+l_2)a}]+ constant
\end{equation}
where $l_1$ and $l_2$ are the lengths of the two intervals, c is the central charge of the $CFT_{1+1}$ and $a$ is the UV cut-off for the $CFT_{1+1}$.

Subsequently the authors addressed the case of a bipartite system in a pure state, described by a single interval in a zero temperature $CFT_{1+1}$. The entanglement negativity for this configuration is obtained by taking the bipartite limit defined by $u_2\to v_1$ and $v_2 \to u_1$ (that is $(A_1,A_2,A^c) \rightarrow(A,A^c,\cancel{0})$) in the four point function given by eq.(\ref{4pt2int}) which leads to the following expression
\begin{equation}\label{0T2pt}
\mathrm{Tr}(\rho_A^{T_2})^{n_e} = 
\langle\overline{{\cal T}}^2_{n_e}(u){\cal T}^2_{n_e}(v)\rangle_{\mathbb{C}}
\end{equation}
where $u_1=v_2$ is denoted as $u$ and $v_2=u_1$ is denoted as $v$.
Since the form of the two point correlation function is fixed completely by the conformal symmetry, the entanglement negativity may be easily computed using eq.(\ref{negrep}) as\footnote{Note that the twist fields ${{\cal T}}^2_{n_e}$ connect $n_e^{th}$ sheet of the Riemann surface $(n_e+2)^{th}$($\overline{{{\cal T}}}^2_{n_e}$ connect $n_e^{th}$ to $(n_e-2)^{th}$) sheet of the Riemann surface. This leads to the factorization of the two point function due to the breaking of $n_e$ even sheeted Riemann surface into two $n_e/2$ sheeted Riemann surfaces.}
\begin{equation}\label{0Ten}
{\cal E}=\frac{c}{2}\log\left(\frac{l}{a}\right)+constant,
\end{equation}
 where $c$ is the central charge, $\ell=|u-v|$ is the length of the subsystem-$A$ and $a$ represents the UV cut-off of the field theory. Interestingly this limit leads to the
expected result from quantum information that for a pure state the entanglement negativity is Renyi entropy of order-$\frac{1}{2}$. 

Although the above procedure works for the pure vacuum state of the $CFT_{1+1}$, eq.(\ref{0T2pt}) is not applicable to the finite temperature mixed state where the $CFT_{1+1}$ is defined on an infinite cylinder \cite{Calabrese:2014yza}. For the latter case the bipartite limit is more subtle and involves the full tripartite system. This configuration involves the subsystems $A, B_1, B_2$ described by the intervals $A=[u_2,v_2]$ of length $\ell$, $B_1=[u_1,v_1]$ and $B_2=[u_3,v_3]$ as depicted in the fig.(\ref{fig3}) below. We denote $B=B_1\cup B_2$ and 
$A^c$ describes the rest of the system.

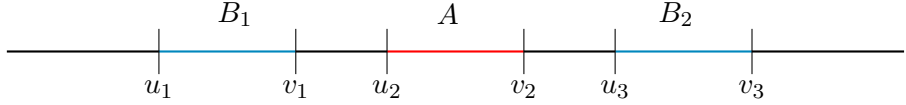
\begin{figure}[H]
  \begin{center}
  \begin{tikzpicture}
  \draw[thick](-5,0)--(-3,0); 
  \draw[thick](-1.2,0)--(-0,0); 
  \draw[thick](1.8,0)--(3,0); 
  \draw[thick](4.8,0)--(6.8,0); 
  \draw[Nblue,thick](-3,0)--(-1.2,0);
  \draw[red,thick ](0,0)--(1.8,0);
  \draw[Nblue,thick](3,0)--(4.8,0);
  \draw(-3,0.3)--(-3,-0.3);
  \draw(-1.2,0.3)--(-1.2,-0.3);
  \draw(0.0,0.3)--(0.0,-0.3);
  \draw(1.8,0.3)--(1.8,-0.3);
  \draw(3.0,0.3)--(3.0,-0.3);
  \draw(4.8,0.3)--(4.8,-0.3);
  \node()at (-2.0,0.5){$B_1$};
  \node()at (-3,-0.5){$u_1$};
  \node()at (-1.2,-0.5){$v_1$};
  \node()at (0.8,0.5){$A$};
  \node()at (3.8,0.5){$B_2$};
  \node()at (0.0,-0.5){$u_2$};
  \node()at (1.8,-0.5){$v_2$};
  \node()at (3.0,-0.5){$u_3$};
  \node()at (4.8,-0.5){$v_3$};
  \end{tikzpicture}
  \end{center}
  \caption{Schematic of the configuration with single interval $A$ in between two disjoint intervals $B_1$ and $B_2$ }
  \label{fig3}
\end{figure}
\begin{figure}[H]
  \begin{center}
  \begin{tikzpicture}
  \draw[thick](-5,0)--(-3,0);
  \draw[Nblue,thick] (-3,0)--(0.0,0); 
  \draw[red,thick] (0,0)--(1.8,0); 
  \draw[Nblue,thick] (1.8,0)--(4.8,0); 
  \draw[thick](4.8,0)--(6.5,0);
  \draw(-3,0.3)--(-3,-0.3);
  \draw(0.0,0.3)--(0.0,-0.3);
  \draw(1.8,0.3)--(1.8,-0.3);
  \draw(4.8,0.3)--(4.8,-0.3);
  \node()at (-1.6,0.5){$B_1$};
  \node()at (-3,-0.5){$u_1$};
  \node()at (0.8,0.5){$A$};
  \node()at (3.4,0.5){$B_2$};
  \node()at (0.0,-0.5){$u_2$};
  \node()at (1.8,-0.5){$v_2$};
  \node()at (4.8,-0.5){$v_3$};
  \end{tikzpicture}
  \caption{Schematic of the configuration in the limit the two intervals $B_1$ and $B_2$ become adjacent to $A$}
  \label{fig4}
  \end{center}
\end{figure}
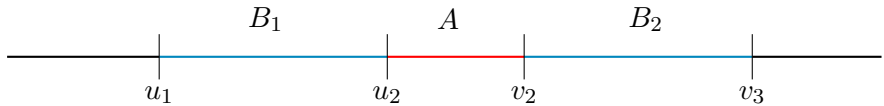

In this case the entanglement negativity is characterized by the six point function of the twist fields as follows
\begin{equation}\label{6pt}
  \mathrm{Tr}(\scaleto{\rho}{8pt}_{\scaleto{AB}{4pt}}^{\scaleto{T_A}{6pt}})^{n_e} = \langle\mathcal{T}_{n_e}(u_1)\overline{\mathcal{T}}_{n_e}(v_1)\overline{\mathcal{T}}_{n_e}(u_2)\mathcal{T}_{n_e}(v_2)\mathcal{T}_{n_e}(u_3)\overline{\mathcal{T}}_{n_e}(v_3)\rangle
 \end{equation}
 In the limit $v_1\to u_2$ and $v_2\to u_3 $ the configuration in fig.(\ref{fig3}) reduced to the one in fig.(\ref{fig4}) and the above six point function reduces to the following four point function
 \begin{equation}\label{4ptvac}
 \mathrm{Tr}(\scaleto{\rho}{8pt}_{\scaleto{AB}{4pt}}^{\scaleto{T_A}{6pt}})^{n_e} = \big<{\cal T}_{n_e}(u_1)\overline{{\cal T}}^2_{n_e}(u_2){\cal T}^2_{n_e}(v_2)\overline{{\cal T}}_{n_e}(v_3)\big>
 \end{equation}
 For the vacuum state of the $CFT_{1+1}$ which lives on the complex plane, the above four point twist correlator has the following form from conformal symmetry \cite{Calabrese:2014yza}
 \begin{equation}\label{rhoAn5}
\big<{\cal T}_{n_e}(z_1)\overline{{\cal T}}^2_{n_e}(z_2){\cal T}^2_{n_e}(z_3)\overline{{\cal T}}_{n_e}(z_4)\big>_{\mathbb{C}}=\frac{c_{n_e}c^2_{n_e/2}}{z_{14}^{2\Delta_{n_e}}z_{23}^{2\Delta^{(2)}_{n_e}}}\frac{{\cal F}_{n_e}(x)}{x^{\Delta^{(2)}_{n_e}}},~~~~~x\equiv\frac{z_{12}z_{34}}{z_{13}z_{24}},
\end{equation}
where $(z_1,z_2,z_3,z_4)=(u_1,u_2,u_3,v_3)$ for the configuration in question. This leads to the following expression for the entanglement negativity of the mixed state configuration depicted in fig.(\ref{fig4})
\begin{equation}\label{Enrmix}
~~~~~~~~~~~{\cal E}=\frac{c}{4}\log \Big(\frac{l_1~l_2^2 ~l_3}{(l_1+l_2)(l_2+l_3)a^2}\Big)+g(x)+ \mathrm{constant},~~~x=\frac{l_1l_3}{(l_1+l_2)(l_2+l_3)}
\end{equation}
where $l_1=|u_1-u_2|$, $l_2=|u_2-v_2|$ and $l_3=|v_2-v_3|$ are the lengths of the intervals $B_1$, $A$ and $B_2$ respectively. The function $g(x)$ and the constant are non universal and depend on the full operator content of the theory. However the end point values of the function $g(x)$ may be fixed to be $g(1)=0$ and $g(0)=const$ as described in \cite{Calabrese:2014yza}.

Interestingly, observe that in the limit  $l_1,l_3>>l_2$, i.e as the cross ratio $x\to1$, the tripartite system involving $A,B$ and the rest of the system reduces to the bipartite system $A\cup A^c$. Note that this limit is equivalent to the bipartite limit $B\to A^c$  leading to $(A\cup B)^c\to \emptyset$. Hence in this limit, the result given by eq.(\ref{Enrmix}) reduces to the following
\begin{equation}\label{Enpu}
~~~~~~~~~~~{\cal E}\approx\frac{c}{2}\log \Big(\frac{l_2}{a}\Big)+ \mathrm{constant}.
\end{equation}
This expression matches exactly with the entanglement negativity of the bipartite pure vacuum state of the $CFT_{1+1}$ given in eq.(\ref{0Ten}). This is expected as in the bipartite limit $B\to A^c$, the mixed state density matrix $\rho_{A\cup B}$ reduces to the pure vacuum state $ \rho_{A\cup A^c}=\ket{0}\bra{0} $ for the case when the full system $A\cup A^c$ is described by the vacuum state of the $CFT_{1+1}$. 

Similarly, as described in \cite{Calabrese:2014yza}, the entanglement negativity of the finite temperature mixed state in the bipartite limit is described through the four point twist correlator as follows
\begin{equation}\label{ENatT}
{\cal E}=\lim_{L \to \infty}\lim_{n_e \to 1}\log\left[\big<{\cal T}_{n_e}(-L)\overline{{\cal T}}^2_{n_e}(-l){\cal T}^2_{n_e}(0)\overline{{\cal T}}_{n_e}(L)\big>_{\mathbb{\beta}}\right],
\end{equation}
Once again, the full tripartite system depicted in fig(\ref{fig4}), reduces to the bipartite system $A\cup A^c$ in the limit $B\to A^c$ ($L\to \infty$), with the coordinates $u_2=-L$ and $v_3=L$ in eq.(\ref{4ptvac}). 
where the subscript $\beta$ indicates that the correlation function has to be evaluated on a cylinder, $v_2=u_1=\ell$ and $v_1=u_3=0$. Note that  the order of the limits is significant to obtain the correct finite temperature negativity, and unlike the zero temperature case discussed previously, the full system $A\cup A^c$ in this case is described by the mixed state thermal density matrix $\rho_{A\cup A^c}=e^{-\beta H}$.

As described by eq.(\ref{ENatT}),  the replica limit $n_e\to1$ has to be imposed prior to the bipartite limit denoted by $L\to \infty$. The four point function of the primary operators is fixed only up to a function of a cross ratios and therefore the entanglement negativity given by eq.(\ref{ENatT}) leads to the following expression
\begin{equation}\label{ENCFTfinite}
{\cal E}=\frac{c}{2}\log\left[\frac{\beta}{\pi a}\sinh\left(\frac{\pi l}{\beta}\right)\right]-\frac{\pi c l}{2\beta}+g(e^{-2\pi l/\beta})+constant.
\end{equation}
The non universal function $g(e^{-2\pi l/\beta})$ and the constant in the above expression depend on the full operator content of the theory. Its values may be fixed only at the end points ($x=0$ and $x=1$). Interestingly the above expression may be expressed as follows
\begin{equation}\label{ENTcon}
 {\cal E}= \frac{3}{2}[S_A-S_A^{th}]+g(e^{-2\pi l/\beta})+constant, 
\end{equation}
where $S_A=\frac{c}{3}\log\left[\frac{\beta}{\pi a}\sinh\left(\frac{\pi l}{\beta}\right)\right]$ and $S_A^{th}=\frac{\pi c l}{3\beta}$ corresponds to the entanglement entropy and the thermal entropy of the subsystem-$A$ respectively. It is clear from the above expression that the  entanglement negativity eliminates the thermal contribution and hence describes the upper bound on the distillable entanglement in a mixed state.
\section{Holographic entanglement negativity conjecture in AdS$_3$/CFT$_2$  }
\begin{figure}[H]
\centering
\includegraphics[scale=2.3]{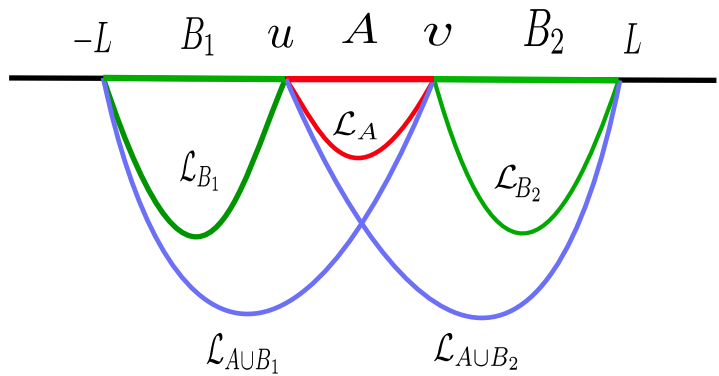}
\caption{\label{henfig} Schematic of geodesics anchored on the subsystems $A$, $B_1$ and $B_2$ in the dual $CFT_{1+1}$.}
\end{figure}

As mentioned in the Introduction it was demonstrated in \cite{Chaturvedi:2016rcn} that the first term describing the universal part of the entanglement negativity admits of a holographic 
description in terms of the bulk geometry which we briefly review in this section. For the mixed state depicted in fig.(\ref{fig4}), the entanglement negativity for the bipartite system $A\cup B$ (where $B=B_1\cup B_2$) was described by a specific algebraic sum of the bulk space like geodesics anchored on appropriate subsystems as shown in fig.(\ref{henfig}) which is given by
\begin{equation}
{\cal E}=\frac{3}{16G_{N}^{(3)}}\bigg[(2 {\cal L}_{A}+{\cal L}_{B_1}+{\cal L}_{B_2}-{\cal L}_{A\cup B_1}-{ \cal L}_{A\cup B_2})\bigg].\label{ENCFT3}
\end{equation}
In the above expression ${\cal L}_{\gamma}$ represents the length of the bulk $AdS_3$ geodesic anchored on the subsystem $\gamma$ in the dual $CFT_{1+1}$. Curiously, upon using the Ryu-Takyanagi conjecture this reduces to the sum of the holographic mutual information between pairs of subsystems in the tripartition as follows
\begin{equation}
{\cal E}=\frac{3}{4}\big[{\cal I}(A,B_1)+{\cal I}(A,B_2)\big].\label{ENCFT4}
\end{equation}
Here, the holographic mutual information between the pair of intervals $(A,B_1)$ and $(A,B_2)$ are given as
\begin{eqnarray}
 {\cal I}(A,B_i)&=& S_{A}+S_{B_i}-S_{A\cup B_i},\nonumber\\
&=&\frac{1}{4G_{N}^{(3)}}({\cal L}_{A}+{ \cal L}_{B_i}-{\cal L}_{A\cup B_i}),~~~~~~\label{MI}
\end{eqnarray}
The bulk dual for the vacuum state of the $CFT_{1+1}$ corresponds to the Pure $AdS_3$ spacetime. The length of bulk geodesic ${\cal L}_{\gamma}$  anchored to the subsystem $\gamma$ in the dual $CFT_{1+1}$ in this space time is given as
\begin{equation}
 {\cal L}_{\gamma}= 2R ~\log\big[\frac{l_{\gamma}}{a}\big].\label{geopu}
\end{equation}
Utilizing  the above eq.(\ref{geopu}) in the holographic conjecture described by eq.(\ref{ENCFT3}), we obtain the following expression for the entanglement negativity for the required mixed state
\begin{equation}\label{Enrlc}
~~~~~~~~~~~{\cal E}=\frac{c}{4}\log \Big(\frac{l_1~l_2^2 ~l_3}{(l_1+l_2)(l_2+l_3)a^2}\Big)
\end{equation}
Notice that remarkably this  matches exactly with the universal part of the entanglement negativity obtained through the replica technique, given by eq.(\ref{Enrmix}). Consider the bipartite limit $B\to A^c$ in which the lengths $l_1,l_3\to \infty$. In this limit the above equation leads to the following result
\begin{equation}\label{Rhalf}
 {\cal E}=\frac{c}{2}\log\left(\frac{l_2}{a}\right)
\end{equation}
 As explained earlier in this limit the mixed state density matrix $\rho_{A\cup B}$ reduces to the pure state $ \rho_{A\cup A^c}=\ket{0}\bra{0}$. 
The above result exactly matches with the universal part of the corresponding replica technique result given in eq.(\ref{Enpu}).

On the other hand for the finite temperature mixed state of a $CFT_{1+1}$, the dual bulk corresponds to the BTZ black hole. In this case the holographic conjecture described by eq.(\ref{ENCFT3}) leads to following expression for entanglement negativity
\begin{eqnarray}\label{ENCFTfinbulk}
{\cal E}&=&\frac{c}{2}\log\left[\frac{\beta}{\pi a}\sinh\left(\frac{\pi l}{\beta}\right)\right]-\frac{\pi c l}{2\beta}\\&=&\frac{3}{2}[S_A-S_A^{th}]
\end{eqnarray}
Once again this matches exactly with the universal part of the corresponding result in eq.(\ref{ENCFTfinite}) obtained through the replica technique. 

The above discussions therefore lead to the significant issue of examining the large central charge limit of the four point twist correlator in eq.(\ref{rhoAn5}) in order to investigate the consistency of this holographic conjecture. In this context, in the subsequent sections, we will utilize the monodromy technique to demonstrate that the universal part of the entanglement negativity given by eq.(\ref{Enrlc}) is dominant in the large central charge limit where as the non universal part of the negativity described by the function $g(x)$ in eq.(\ref{Enrmix}) and eq.(\ref{ENTcon}) is sub leading. This would serve as a stringent consistency check from the $CFT_{1+1}$ for the holographic entanglement negativity  conjecture described in \cite {Chaturvedi:2016rcn,Chaturvedi:2016rft,Chaturvedi:2016opa}. We proceed to address this issue in the following sections.

\section{ Conformal blocks at large-c and the Monodromy technique}

In this section we will review the monodromy technique for examining the large-c behavior of the conformal blocks of  four point correlation functions in a $CFT_{1+1}$\cite{Hartman:2013mia,Fitzpatrick:2014vua}. To this end the authors in \cite{Hartman:2013mia,Fitzpatrick:2014vua} consider the well known s-channel ($x\to0$) conformal block expansion of the four point function of the primary operators ${\cal O}_i$ in a $CFT_{1+1}$
which is given by 
\begin{equation}\label{4pcb}
\langle{\cal O}_1(0){\cal O}_2(x){\cal O}_3(1){\cal O}_4(\infty)\rangle=\sum_{p}C_{12}^{p}C_{34}^{p}{\cal F}(x,h_p,h_i)\overline{{\cal F}}(\overline{x},\overline{h}_p,\overline{h}_i) .
\end{equation}
In the above expression the sum is over all the primary operators ${\cal O}_p$ with scaling dimensions $h_p$, $C_{12}^{p},C_{34}^{p}$ are the three point OPE coefficients and the functions  
${\cal F}(x,h_p,h_i)$ represent the corresponding conformal blocks. Note that the four point function in eq.(\ref{4pcb}) is in conformally transformed coordinates given by
$(0,x,1,\infty)$ and may be obtained from the original four point function as follows
\begin{equation}
 \langle{\cal O}_1(0){\cal O}_2(x){\cal O}_3(1){\cal O}_4(\infty)\rangle=\lim_{z_4,\overline{z}_4\to\infty}z_4^{2h_4}\overline{z}_4^{2\overline{h}_4}\langle{\cal O}_1(0){\cal O}_2(x){\cal O}_3(1){\cal O}_4(z_4)\rangle.
\end{equation}

In \cite{Zamolodchikov:1995aa}  Zamolodchikov et al. demonstrated  that in the limit $c\to\infty$ the conformal blocks of the four point function of the primary operators in a Liouville field theory exponentiate as follows (see \cite{Harlow:2011ny} for a review)
\begin{equation}\label{cbexp}
\mathcal{F}(c,h_p,h_i,x) \sim\exp\Big[-\frac{c}{6}f\Big(\frac{h_p}{c},\frac{h_i}{c},x\Big)\Big].
\end{equation}
In the above mentioned approximation it is assumed that $\frac{h_p}{c}$ and $\frac{h_i}{c}$ are held fixed in the large-c limit. Although a general proof of this statement remains elusive there is strong indication that the above mentioned exponentiation of the conformal blocks is true for all holographic $ CFT_{1+1}$s \cite{Zamolodchikov1987, Zamolodchikov1984,Fitzpatrick:2014vua}. The function 
$f\Big(\frac{h_p}{c},\frac{h_i}{c},x\Big)$ is obtained by determining the monodromy of the solutions of a certain second order differential equation which is given as follows
\begin{equation}\label{ODE}
\psi''(z) + T(z)\psi(z) = 0. 
\end{equation}
The function $\psi(z)$ is obtained by inserting a light degenerate operator $\hat{\psi}(z)$ ($ h_{\psi}\sim O[1]$) in the four point function of the above mentioned operators 
${\cal O}_i$,  which in the $c\to \infty$ limit leads to the following simplification
\begin{equation}\label{psiz}
\sum_{p}\langle{\cal O}_1(z_1){\cal O}_2(z_2)\ket{{\cal O}_p}\bra{{\cal O}_p}\hat{\psi}(z){\cal O}_3(z_3){\cal O}_4(z_4)\rangle\approx \psi(z,z_i) \sum_{p}\langle{\cal O}_1(z_1){\cal O}_2(z_2)\ket{{\cal O}_p}\bra{{\cal O}_p}{\cal O}_3(z_3){\cal O}_4(z_4)\rangle.
\end{equation}
On the other hand the function $T(z)$ in eq.(\ref{ODE}) is obtained by using eq.(\ref{cbexp}), eq.(\ref{psiz}) and the conformal ward identity as follows
\begin{eqnarray}
\langle{\hat{T}(z)\cal O}_1(z_1){\cal O}_2(z_2){\cal O}_3(z_3){\cal O}_4(z_4)\rangle&=&-\frac{c}{6}T(z)\langle{\cal O}_1(z_1){\cal O}_2(z_2){\cal O}_3(z_3){\cal O}_4(z_4)\rangle\label{5ptT}\\
T(z)&=&\sum_{i=1}^{4}\frac{h_i}{(z-z_i)^2}+\frac{c_i}{(z-z_i)}.\label{stress}
\end{eqnarray}
Three of these parameters $c_1,c_3$ and $c_4$ in eq.(\ref{stress}) are fixed by the constraint $T(z)\sim z^{-4}$ as $z\to\infty$ and may be expressed as
\begin{equation}\label{cis}
\sum_i c_i=0 \hspace{1cm} \sum_{i}( c_i z_i- \frac{6 h_i}{c})=0 \hspace{1cm}  \sum_{i} (c_i z_i^2- \frac{12 h_i }{c}z_i)=0.
\end{equation}
The fourth accessory parameter $c_2$ is given by the derivatives of the required function $f(x)$ as $c_2=\frac{\partial f}{\partial x}$.
A fourth constraint is imposed using OPEs in corresponding channels to determine this accessory parameter $c_2$  by fixing the trace of the monodromy matrix around the singularities of $T(z)$ to be as follows (See \cite{Fitzpatrick:2014vua} for details)
\begin{equation}\label{monodromy}
 Tr(\textbf{M}_{0x})=-2\cos{(\pi\Lambda_p)},\hspace{0.5cm} h_p=\frac{c}{24}(1-\Lambda_p^2).
\end{equation}
Here $h_p$ is the scaling dimension of the intermediate operator ${\cal O}_p$ in the given channel with the lowest scaling dimension. Hence, the above mentioned monodromy condition leads to the fourth accessory parameter $c_2=\frac{\partial f}{\partial x}$.  The required function $f(x)$ may then be obtained upon integration.

\subsection{The two adjacent interval case}\label{secadj}
To illustrate the monodromy technique described above we briefly outline its application to compute the entanglement negativity for the mixed state of the two adjacent intervals in the holographic $CFT_{1+1}$  as described in \cite{Kulaxizi:2014nma}. There the authors utilize the above technique to determine the conformal block for the four point twist correlator given by eq.(\ref{4pt2int}) which characterizes the entanglement negativity of the mixed state in question. Note that the case of adjacent interval corresponds to the $t$-channel for the four point correlator described by  the limit $x\to1$. Hence, in this limit the four point function in eq.(\ref{4pt2int}) reduces to the three point function as described by eq.(\ref{3ptadj}). In this case the authors in \cite{Kulaxizi:2014nma} demonstrated that the function $T(z)$ in eq.(\ref{stress}) is given by
\begin{equation}\label{Tadj}
T(z)\approx \frac{c_2(1-x)}{(z-1)^2}.
\end{equation}
Therefore the second order ordinary differential equation given in eq.(\ref{ODE}) reduces to the following 
\begin{equation}\label{ODEadj}
\psi''(z) + \frac{c_2(1-x)}{(z-1)^2}\psi(z) = 0 .
\end{equation}
Hence the monodromy condition in eq.(\ref{monodromy}) imposed on the solutions of this differential equation leads to the following expression for the function $f(x)$
\begin{equation}\label{pfadj}
\frac{\partial f}{\partial x}=-\frac{3}{4(1-x)}.
\end{equation}
Upon integration the above partial differential equation leads to the entanglement negativity of the adjacent intervals in the large central charge limit as follows \cite{Kulaxizi:2014nma}
\begin{equation}\label{fadj}
 f(x)=\frac{3}{4} \log [(1-x)]
\end{equation}
As described in \cite{Calabrese:2012nk,Kulaxizi:2014nma}, in the limit $x\to1$ which corresponds to $u_2\to v_1$, denoting $\epsilon=u_2-v_1$ with $l_j=v_j-u_j~ \{j=1,2\}$, the above expression for the function $f(x)$, the large central charge limit of the entanglement negativity reduces to the following 
\begin{equation}\label{enlcadj}
{\cal E}=\frac{c}{4} \log [\frac{l_1l_2}{(l_1+l_2)\epsilon}]+O[\epsilon].
\end{equation}
Interestingly upon setting $\epsilon=a$, their result matches exactly with the universal part of the entanglement negativity in the $CFT_{1+1}$ obtained through the replica technique in eq.(\ref{enadj1})\cite{Calabrese:2012ew,Calabrese:2012nk}. This naturally implies that the non-universal constant in eq.(\ref{enadj1}) is subleading in the large central charge limit. The authors also numerically demonstrated that in the s-channel described by the limit $x\to0$ for the case when the two intervals are disjoint and far apart the entanglement negativity vanishes. Motivated by these results the present authors in collaboration have recently proposed a holographic conjecture for the entanglement negativity of mixed states described by such adjacent subsystems in holographic CFTs \cite{Jain:2017aqk,Jain:2017xsu}. This conjecture involves a specific algebraic sum of the areas of extremal surfaces in the bulk geometry which reduces to the holographic mutual information between the subsystems upto a numerical factor. Remarkably it could be shown that for the $AdS_3/CFT_2$ case the holographic entanglement negativity computed from this conjecture exactly reproduces the result given by eq.(\ref{enlcadj}) in \cite{Kulaxizi:2014nma}.

\section{Four point function in the large central charge limit}\label{fopt}
In this section we now focus our attention to the large central charge analysis of the four point twist correlator in eq.(\ref{ENatT}) which characterizes the entanglement negativity of the mixed state configuration depicted in fig. (\ref {fig4}). The form of this four point twist correlator may be fixed up to a function of the cross ratios as follows\cite{Calabrese:2014yza}
\begin{equation}\label{4ptform}
\big<{\cal T}_{n_e}(z_1)\overline{{\cal T}}^2_{n_e}(z_2){\cal T}^2_{n_e}(z_3)\overline{{\cal T}}_{n_e}(z_4)\big>_{\mathbb{C}}=\frac{c_{n_e}c^{(2)}_{n_e}}{z_{14}^{2\Delta_{n_e}}z_{23}^{2\Delta^{(2)}_{n_e}}}\frac{{\cal F}_{n_e}(x)}{x^{\Delta^{(2)}_{n_e}}},~~~~~x\equiv\frac{z_{12}z_{34}}{z_{13}z_{24}}.
\end{equation}
where $x$ is the cross ratio, the intervals $z_{ij}=z_i-z_j$ and the function ${\cal F}_{n_e}(x)$ depends on full operator content of the theory.  We now implement the following conformal transformation to map the points $(z_1,z_2,z_3,z_4)\to(0,x,1,\infty$)
\begin{equation}\label{cmap}
 w_i=\frac{(z_i-z_1)(z_3-z_4)}{(z_i-z_4)(z_3-z_1)}.
\end{equation}
The four point function in eq.(\ref{4ptform}) then transforms as follows \cite{francesco1996conformal}
\begin{eqnarray}
 \langle{\cal T}_{n_e}(0)\overline{{\cal T}}^2_{n_e}(x){\cal T}^2_{n_e}(1)\overline{{\cal T}}_{n_e}(\infty)\rangle
&=& \lim_{z_4\to\infty}|z_4|^{2\Delta_4}\langle{\cal T}_{n_e}(0)\overline{{\cal T}}^2_{n_e}(x){\cal T}^2_{n_e}(1)\overline{{\cal T}}_{n_e}(z_4)\rangle \\
 &=& \frac{c_{n_e}c^{(2)}_{n_e}}{(1-x)^{2\Delta^{(2)}_{n_e}}}\frac{{\cal F}_{n_e}(x)}{x^{(\Delta^{(2)}_{n_e}+\Delta_{n_e})}}\label{4pttran}.
\end{eqnarray}

We will now proceed to obtain the large central charge limit of the above four point twist correlator in different channels using the monodromy technique. Note that the operator ${\cal T}_{n_e}$ and $\overline{{\cal T}}_{n_e}$ have the scaling dimension $h_{{\cal T}_{n_e}}=\frac{c}{24}(n_e-\frac{1}{n_e})$ which in the replica limit $n_e\to 1$ vanishes ($h_{{\cal T}}\to0$). Therefore these are light operators in the large central charge limit. On the other hand the operators ${\cal T}^2_{n_e}$ and $\overline{{\cal T}}_{n_e}^2$ in the four point function given by eq.(\ref{4ptform}) have the scaling dimension $h_{{\cal T}_{n_e}^2}=2h_{{\cal T}_{\frac{n_e}{2}}}=\frac{c}{12}(\frac{n_e}{2}-\frac{2}{n_e})$ as explained in the previous chapter. Therefore in the replica limit $n_e\to1$, we have $h_{{\cal T}^2}\to -\frac{c}{8}$ making these operators heavy in the large central charge limit as
$O[h_{{\cal T}^2}]\sim c$. We now denote $\frac{h_{{\cal T}}}{c}$ as $\epsilon_L$ and $\frac{h_{{\cal T}^2}}{c}$ as $\epsilon_H$ so the function $T(z)$ in eq.(\ref{stress}) may be expressed as 
\begin{equation}\label{T(z)}
 T(z)=\frac{6 \epsilon _H}{(z-x)^2}-\frac{12 \epsilon _H}{z-1}+\frac{6 \epsilon _H}{(z-1)^2}+\frac{12 \epsilon _H}{z}+\frac{6 \epsilon _L}{z^2}-\frac{x(  x-1)c_2}{(z-x) (z-1) z}.
\end{equation}

In order to fix the monodromy condition given in eq.(\ref{monodromy}) we require the dimensions of the intermediate operators in different channels. These may be obtained through the OPEs of ${\cal T}_{n_e}(z_i)$ with $\overline{{\cal T}}^2_{n_e}(z_j)$ and $\overline{{\cal T}}^2_{n_e}(z_i)$ with ${\cal T}^2_{n_e}(z_j)$ as $z_i\to z_j$ that are given as follows \cite{Calabrese:2014yza}
\begin{eqnarray}\label{tauOPEs}
 {\cal T}_{n_e}(z_i)\overline{{\cal T}}_{n_e}^2(z_j)&=& \frac{C_{{\cal T}_{n_e}\overline{{\cal T}}_{n_e}^2\overline{{\cal T}}_{n_e}}}{(z_i-z_j)^{2\Delta_{n_e}}}\overline{{\cal T}}_{n_e}(z_j)+.....\\
 \overline{{\cal T}}_{n_e}^2(z_i){\cal T}_{n_e}^2(z_j)&=& \frac{c_{n_e}^{(2)}}{(z_i-z_j)^{2\Delta_{n_e}^{(2)}}}{\cal I}+.....
\end{eqnarray}
As explained in\cite{Kulaxizi:2014nma}, in the large central charge limit, the leading contribution for the correlation function corresponding to the negativity comes from the conformal block of the intermediate operator with the lowest scaling dimension obtained by fusing the twist operators in their respective channels. Hence, from the above OPEs it is clear that the leading contribution for the required four point function comes from the intermediate operators $\overline{{\cal T}}_{n_e}$ and the identity operator ${\cal I}$ in the $s$ and $t$ channels respectively. 

\subsection{t-channel ( pure state limit)}
We now proceed to determine the conformal block which provides the dominant contribution in the large central charge limit, for the four point correlator in question in the $t$-channel ($x\to 1$). Note that in this limit  the length of the intervals implies $l_1,l_3>>l_2$. As explained earlier, in this limit
 the required mixed state configuration which is depicted in fig.(\ref{fig4}) reduces to that of a pure state. To this end we note that the leading singular part of the function $T(z)$ near $z=x=1$ is given as 
\begin{equation}\label{TzinT}
 T(z)=\frac{12 \epsilon_H-c_2(1-x)}{(z-1)^2}.
\end{equation}
Hence the solution to the differential equation given by eq.(\ref{ODE}) with $T(z)$ as above, is 
\begin{equation}\label{psiinT}
  \psi_{\pm}=(z-1)^{\frac{1}{2}(1\pm\sqrt{1-4 (12\epsilon_H-c_2(1-x)})}.
\end{equation}
The monodromy condition given in eq.(\ref{ODE}) leads to the following expression for the parameter $c_2(x)$ as follows
\begin{equation}\label{c2inT}
 c_2(x)=\frac{\partial f}{\partial x}=\frac{12 \epsilon_H}{(1-x)}.
\end{equation}
Hence, upon integration we obtain the required function $f(x)$ as
\begin{equation}\label{finT}
 f(x)= 12 \epsilon_H \log[1-x]+O[\epsilon_H^2].
\end{equation}
 Exponentiating this function to obtain the conformal block as described by eq.(\ref{cbexp}) leads to the following expression for the four point correlator in the large-c limit as
\begin{equation}\label{4ptinT}
  \langle{\cal T}_{n_e}(0)\overline{{\cal T}}^2_{n_e}(x){\cal T}^2_{n_e}(1)\overline{{\cal T}}_{n_e}(\infty)\rangle\approx \frac{1}{(1-x)^{4c\epsilon_H}}=\frac{1}{(1-x)^{2\Delta^{(2)}_{n_e}}}.
 \end{equation}
  From eq.(\ref{4ptform}) and eq.(\ref{4pttran}), note that the above four point function is related to the four point function on complex plane as follows
 \begin{equation}
 \big<{\cal T}_{n_e}(z_1)\overline{{\cal T}}^2_{n_e}(z_2){\cal T}^2_{n_e}(z_3)\overline{{\cal T}}_{n_e}(z_4)\big>_{\mathbb{C}}=\frac{(1-x)^{2\Delta^{(2)}_{n_e}}x^{\Delta_{n_e}}}{z_{14}^{2\Delta_{n_e}}z_{23}^{2\Delta^{(2)}_{n_e}}}\langle{\cal T}_{n_e}(0)\overline{{\cal T}}^2_{n_e}(x){\cal T}^2_{n_e}(1)\overline{{\cal T}}_{n_e}(\infty)\label{transf}
 \end{equation}
 Upon substituting the result given by eq.(\ref{4ptinT}) in  eq.(\ref{transf}), we obtain to the following expression for the large central charge limit of the required four point twist correlator in $t$-channel
 \begin{equation}
  \big<{\cal T}_{n_e}(z_1)\overline{{\cal T}}^2_{n_e}(z_2){\cal T}^2_{n_e}(z_3)\overline{{\cal T}}_{n_e}(z_4)\big>_{\mathbb{C}}\approx\frac{1}{z_{14}^{2\Delta_{n_e}}z_{23}^{2\Delta^{(2)}_{n_e}}}
 \end{equation}
Hence the above result leads to the following expression for the entanglement negativity in the $t$-channel in the large central charge limit
\begin{equation}
 {\cal E}=\frac{c}{2}\log[\frac{l_2}{a}]
\end{equation}
Interestingly the above large central charge result matches exactly with the  eq.(\ref{Rhalf}) obtained using the holographic conjecture described in \cite{Chaturvedi:2016rcn}, in the limit $l_1,l_3>>l_2$ which here corresponds to $t$-channel ($x\to1$). As described earlier in this limit $l_1,l_3>>l_2$ the mixed state configuration in question reduces to that of the pure state and hence, the above result also matches with the universal result given by eq.(\ref{Enpu}) which was obtained through the replica technique in \cite{Calabrese:2012ew,Calabrese:2012nk,Calabrese:2014yza}.  In the next section we will examine the large central charge limit of the required four point twist correlator corresponding to the entanglement negativity of the mixed state in question, in the $s$-channel.

\subsection{s-channel}
Having examined the four point correlator in question in the $t$-channel, we now proceed to investigate its large central charge behaviour in the $s$-channel which is described by the limit $x\to0$. To this end we note that the leading singular part of the function $T(z)$ in eq.(\ref{stress}) near $z=x=0$ is given by
\begin{equation}\label{TinS}
 T(z)=\frac{6\epsilon_H-c_2 x}{z^2}.
\end{equation}
Hence, the solution to the differential equation given by eq.(\ref{ODE}) for this $T(z)$ is
\begin{equation}\label{psiins}
 \psi_{\pm}=z^{\frac{1}{2}(1\pm\sqrt{1-4(6 \epsilon_H-c_2 x)})}.
\end{equation}
Using the monodromy condition given by eq.(\ref{monodromy}) we obtain following expression for the accessory parameter $c_2$
\begin{equation}\label{c2ins}
 c_2=\frac{\partial f}{\partial x}=\frac{6 \epsilon_H}{x}.
\end{equation}
Upon integrating the above expression we find the required function $f(x)$ as follows
 \begin{equation}\label{fins}
  f(x)=6\epsilon_H\log[x]+O[\epsilon_H^2].
 \end{equation}
Note that the conformal blocks obey a recursive relation which was used to determine it as an expansion in $x$ near $x=0$ given in \cite{Hartman:2013mia}
 \begin{equation}\label{recur}
  \frac{c}{6}f(x)= (h_1+h_2-h_p )\log[x]+O[x].
 \end{equation} 
 For the required s-channel conformal block we have $\frac{h_1}{c}=\epsilon_L=0$, $\frac{h_2}{c}=\epsilon_H$ and $h_p=0$ which leads to the following expression
\begin{equation}
 f(x)=6\epsilon_H \log[x]+O[x].
\end{equation}
This matches exactly with the result we obtained in eq.(\ref{fins}).
Finally, exponentiating this function leads to the following expression for the required four point correlator in the large central charge limit for the $s$-channel as follows
 \begin{equation}\label{4ptins}
  \langle{\cal T}_{n_e}(0)\overline{{\cal T}}^2_{n_e}(x){\cal T}^2_{n_e}(1)\overline{{\cal T}}_{n_e}(\infty)\rangle\approx \frac{1}{x^{2c\epsilon_H}}=\frac{1}{x^{\Delta^{(2)}_{n_e}} }.
 \end{equation}
Once again utilizing the above equation in the transformation given by eq.(\ref{transf}) we obtain the twist correlators in original $z_i$ coordinates as
\begin{equation}
 \big<{\cal T}_{n_e}(z_1)\overline{{\cal T}}^2_{n_e}(z_2){\cal T}^2_{n_e}(z_3)\overline{{\cal T}}_{n_e}(z_4)\big>_{\mathbb{C}}\approx\frac{x^{\Delta_{n_e}}}{x^{\Delta^{(2)}_{n_e}} z_{14}^{2\Delta_{n_e}}z_{23}^{2\Delta^{(2)}_{n_e}}}
\end{equation}
On identifying the coordinates $(z_1,z_2,z_3,z_4)=(u_1,v_1,u_2,v_2)$ with $|u_1-v_1|=l_1$,$|v_2-u_1|=l_2$ and $|u_2-v_2|=l_3$, the above equation leads to the following expression for the entanglement negativity in this channel 
\begin{equation}\label{enins}
{\cal E}=\frac{c}{4}\log \Big(\frac{l_1~l_2^2 ~l_3}{(l_1+l_2)(l_2+l_3)a^2}\Big).
\end{equation}
Quite interestingly, this matches exactly with the corresponding result obtained using the holographic negativity conjecture given in eq.(\ref{ENCFT3}). This leads us to conclude that the function $g(x)$ in the expression for the full entanglement negativity given by eq.(\ref{Enrmix}) is sub leading in the large central charge limit in both the $s$ and the $t$ channels\footnote{Note that it might  seem as if there could be phase transition of the entanglement negativity between its values in $t$ and $s$- channel. This is not the case as the phase transitions for entanglement entropy in the large central charge limit reported in \cite{Hartman:2013mia} and for entanglement negativity in \cite{Kulaxizi:2014nma} are between their values in $t$-channel (when the intervals are close by) and $s$-channel (when the intervals are far apart). We emphasize here that this is not the situation for the mixed state configuration considered depicted in fig.(\ref{fig4}), as the intervals $B_1$ and $B_2$ are always adjacent to the interval $A$. This is also clear from the fact that this four point function arises from the fusion of twist operators in a six point correlator in a specific limit as will be described in section.(\ref{sixpo}).}.

 The computations in this section therefore provide clear evidence that the function $g(x)$ is subleading in the large $c$ limit, in both $s$ and $t$ channels (i.e as $x\to 0$ and  $x\to 1$). However, in order to determine the behavior of entanglement negativity everywhere between $0$ and $1$ ($0\leq x \leq1$) we need to begin by considering the six point function in eq.(\ref{6pt}) instead of the four point function given by eq.(\ref{4ptform}) in a specific limit. We proceed to perform this computation in the next section.

\section{Six point function in the large central charge limit }\label{sixpo}

\begin{figure}[H]
  \begin{center}
  \begin{tikzpicture}
  \draw[thick](-5,0)--(-3,0); 
  \draw[thick](-1.2,0)--(-0,0); 
  \draw[thick](1.8,0)--(2.7,0); 
  \draw[thick](4.8,0)--(6.8,0); 
  \draw[Nblue,thick](-3,0)--(-0.9,0);
  \draw[red,thick ](0,0)--(1.8,0);
  \draw[Nblue,thick](2.7,0)--(4.8,0);
  \draw(-3,0.3)--(-3,-0.3);
  \draw(-0.9,0.3)--(-0.9,-0.3);
  \draw(0.0,0.3)--(0.0,-0.3);
  \draw(1.8,0.3)--(1.8,-0.3);
  \draw(2.7,0.3)--(2.7,-0.3);
  \draw(4.8,0.3)--(4.8,-0.3);
  \node()at (-2.0,0.5){$B_1$};
  \node()at (-3,-0.5){$z_1$};
  \node()at (-0.9,-0.5){$z_2$};
  \node()at (0.8,0.5){$A$};
  \node()at (3.8,0.5){$B_2$};
  \node()at (0.0,-0.5){$z_3$};
  \node()at (1.8,-0.5){$z_4$};
  \node()at (2.7,-0.5){$z_5$};
  \node()at (4.8,-0.5){$z_6$};
  \end{tikzpicture}
  \end{center}
  \caption{Schematic of the configuration with single interval $A$ in between two disjoint intervals $B_1$ and $B_2$ }
  \label{fig5}
\end{figure}
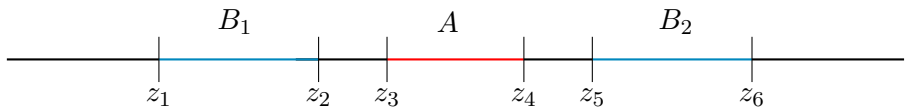
In this section we examine the large central charge limit of the entanglement negativity for the bipartite mixed state  described by a single interval ($A$) with two intervals ($B_1\cup B_2=B$) on either side as depicted in fig.(\ref{fig5}). We will utilize the monodromy technique to obtain the large-$c$ behavior of the six point function corresponding to the entanglement negativity, in the limit in which the two intervals $B_1$ and $B_2$ become adjacent to $A$.  Hence, in order to determine the entanglement negativity for this mixed state it is required to compute the following six point function  in the limit $z_2\to z_3$ and $z_4\to z_5$
\begin{equation}\label{6ptz}
  \mathrm{Tr}(\scaleto{\rho}{8pt}_{\scaleto{AB}{4pt}}^{\scaleto{T_A}{6pt}})^{n_e} = \langle\mathcal{T}_{n_e}(z_1)\overline{\mathcal{T}}_{n_e}(z_2)\overline{\mathcal{T}}_{n_e}(z_3)\mathcal{T}_{n_e}(z_4)\mathcal{T}_{n_e}(z_5)\overline{\mathcal{T}}_{n_e}(z_6)\rangle.
 \end{equation}
A generic six point function in a $CFT_{1+1}$ depends on three independent conformal cross ratios which we choose to be
\begin{equation}\label{cr}
 x_1=\frac{z_{21}z_{56}}{z_{26}z_{51}},\ \ x_2=\frac{z_{31}z_{56}}{z_{36}z_{51}}\ \ and \ \ x_3=\frac{z_{41}z_{56}}{z_{46}z_{51}}
\end{equation}
where $z_{ij}=z_i-z_j$ such that the points $(z_1,z_2,z_3,z_4,z_5,z_6)$ are situated as shown in fig.(\ref{fig5}) and the lengths of the intervals $B_1$, $A$ and $B_2$ are given as ($l_1=|z_1-z_2|, l_2=|z_3-z_4|, l_3=|z_5-z_6|)$. We now consider a conformal transformation $w=\frac{(z-z_1)(z_5-z_6)}{(z-z_6)(z_5-z_1)}$ such that the six point function in eq.(\ref{6ptz}) depends only on the above mentioned cross ratios $x_1,x_2$ and $x_3$ as follows
\begin{eqnarray}
  G_{n_e}(x_1,x_2,x_3)&=&\langle{\cal T}_{n_e}(0)\overline{{\cal T}}_{n_e}(x_1)\overline{{\cal T}}_{n_e}(x_2){\cal T}_{n_e}(x_3){\cal T}_{n_e}(1)\overline{{\cal T}}_{n_e}(\infty)\rangle\label{sixG} \\
&=& \lim_{z_6\to\infty}|z_6|^{2\Delta_6}\langle{\cal T}_{n_e}(0)\overline{{\cal T}}_{n_e}(x_1)\overline{{\cal T}}_{n_e}(x_2){\cal T}_{n_e}(x_3){\cal T}_{n_e}(1)\overline{{\cal T}}_{n_e}(z_6)\rangle \nonumber
\end{eqnarray}
Note that the required limits $z_2\to z_3$ and $z_4\to z_5$ as depicted in fig.(\ref{fig5}) correspond to the channel  defined by the limits $x_1\to x_2$ and $x_3\to 1$ in the transformed coordinates. The entanglement negativity for the  mixed state configuration in question, is therefore given by
\begin{equation}\label{entG6}
 {\cal E}= \lim_{n_e\to1} \log[G_{n_e}(x_1,x_2,x_3)_{\Omega_1}]
\end{equation}
where the subscript $\Omega_1$ denotes the limits $x_1\to x_2$ and $x_3\to 1$. 
In the large central charge limit once again the dominant conformal block is expected to exponentiate as follows
\begin{equation}\label{G6exp}
 G_{n_e}(x_1,x_2,x_3)\approx \exp{\big[-\frac{c}{6}f\big(\epsilon_p,\epsilon_i, x_1,x_2,x_3\big)}\big]
\end{equation}
where $\epsilon_p=\frac{h_p}{c}$ and $\epsilon_i=\frac{h_i}{c}$ and the function 
$f\big(\epsilon_p,\epsilon_i, x_1,x_2,x_3\big)$ has to be determined from the monodromies around the singularities of the stress tensor $T(z)$ which is given by
\begin{equation}\label{T6}
 T(z)=\sum_{i=1}^{6}\frac{h_i}{(z-z_i)^2}+\frac{c_i}{(z-z_i)}.
\end{equation}
As discussed for the case of the four point function in section.(\ref{fopt}), three of these parameters are fixed by demanding that the stress tensor $T(z)\sim z^{-4}$ for large $z$ which leads to the following constraints on the accessory parameters $c_i$ 
\begin{equation}\label{cis6}
\sum_{i=1}^6 c_i=0 \hspace{1cm} \sum_{i=1}^{6}( c_i z_i- \frac{6 h_i}{c})=0 \hspace{1cm}  \sum_{i=1}^6 (c_i z_i^2- \frac{12 h_i }{c}z_i)=0.
\end{equation}
Identifying the required points as $(z_1,z_2,z_3,z_4,z_5,z_6)=(0,x_1,x_2,x_3,1,\infty)$ and then substituting the above equations in the expression for the stress tensor $T(z)$ in eq.(\ref{cis6}) we arrive at the following expression
\begin{eqnarray}\label{stress6}
 T(z)&=& \frac{6 h}{c z^2}+\frac{6 h}{c (z-1)^2}+\frac{6 h}{c \left(z-x_1\right){}^2}+\frac{6 h}{c \left(z-x_2\right){}^2}+\frac{6 h}{c \left(z-x_3\right){}^2}-\frac{c_2}{z-x_1}-\frac{c_3}{z-x_2}-\frac{c_4}{z-x_3}\nonumber\\&&-\frac{\frac{24 h}{c}-c_2 x_1-c_3 x_2-c_4 x_3}{z-1}-\frac{c_2 \left(x_1-1\right)+ c_3 \left(x_2-1\right)+ c_4 x_3- c_4-24 \frac{h}{c}}{ z}
\end{eqnarray}
Note that since we have used the three constraints in eq.(\ref{cis6}), the stress tensor $T(z)$ now depends only on the three accessory parameters $c_2,c_3$ and $c_4$ which are related to the derivatives of the function $f\big(\epsilon_p,\epsilon_i, x_1,x_2,x_3\big)$ as follows
\begin{equation}\label{acess}
 c_2=\frac{\partial f}{\partial x_1},~~ c_3=\frac{\partial f}{\partial x_2}~~ \mathrm
{and} ~~c_4=\frac{\partial f}{\partial x_3}
\end{equation}
 The trace of the monodromies around the singularities of the stress tensor in a given channel are expected to obey the following condition
\begin{equation}\label{mono61}
  Tr(\textbf{M})=-2\cos{(\pi\Lambda_p)},\hspace{0.5cm} h_p=\frac{c}{24}(1-\Lambda_p^2).
\end{equation}

As described in \cite{Kulaxizi:2014nma} for the case of the entanglement negativity for adjacent intervals, the conformal blocks corresponding to the intermediate operators ($\overline{{\cal T}}_{n_e}^2$) provide the dominant contribution in the large central charge limit. This is because for the entanglement negativity, the operators that fuse in the limit of intervals becoming adjacent are two identical twist operators. Note that this is different from the case for entanglement entropy of multiple intervals considered in \cite{Hartman:2013mia}, where a twist and an anti-twist operator fuse in pairs, and the vacuum block provides the dominant contribution. As is evident from the ordering of the twist operators in eq.(\ref{sixG}), the intermediate operators for the six point correlator in the limits $x_1\to x_2$ and $x_3\to 1$  are also $\overline{{\cal T}}_{n_e}^2$ and ${\cal T}_{n_e}^2$ respectively as in \cite{Kulaxizi:2014nma}. Upon using the scaling dimension of the operators $\overline{{\cal T}}_{n_e}^2$ and ${\cal T}_{n_e}^2$ given by $h_p=-\frac{c}{8}$ in the replica limit, eq.(\ref{mono61}) reduces to
\begin{equation}\label{mono6}
 Tr(\textbf{M}_{x_1 x_2})= Tr(\textbf{M}_{x_3 1})=-2
\end{equation} 
The stress tensor $T(z)$ near $z=x_1= x_2$ in the replica limit 
$h\to 0$ is given as follows
\begin{equation}\label{Tx1x2}
 T(z)\approx-\frac{c_2 \left(x_1-x_2\right)}{\left(z-x_2\right)^2}
\end{equation}
The solutions for the differential equation given by eq.(\ref{ODE}) with the above expression for $T(z)$ are as follows
\begin{equation}\label{si6}
  \psi_{\pm}=(z-x_2)^{\frac{1}{2}(1\pm\sqrt{1-4c_2(x_2-x_1)}}.
\end{equation}
The monodromy condition given in eq.(\ref{mono6}) for the above solutions, leads to the following expression for the accessory parameter $c_2$
\begin{equation}\label{acc6x1x2}
 c_2=\frac{\partial f}{\partial x_1}=\frac{3}{4(x_1-x_2)}
\end{equation}

Similarly the stress tensor $T(z)$ near $z=x_3= 1$ is given as follows
\begin{equation}\label{Tx31}
 T(z)\approx\frac{c_4(1- x_3)}{(z-1)^2}
\end{equation}
The solutions for the differential equation in eq.(\ref{ODE}) with the above expression for $T(z)$ are given as
\begin{equation}\label{si62}
  \psi_{\pm}=(z-1)^{\frac{1}{2}(1\pm\sqrt{1-4 (c_4(1-x_3)})}.
\end{equation}
As earlier, upon using the conditions on the trace of the monodromy matrix given in eq.(\ref{mono6}) for the above solutions we arrive at
\begin{equation}\label{acc6x31}
 c_4=\frac{\partial f}{\partial x_3}=\frac{3}{4(x_3-1)}
\end{equation}

Upon integrating eq.(\ref{acc6x1x2}) and eq.(\ref{acc6x31}), it may be shown that the form of the required function $f(x_1,x_2,x_3)$ is  fixed upto a function of $x_2$ which we denote as $\phi(x_2)$, to be as follows
\begin{equation}\label{fullf}
 f(x_1,x_2,x_3)=\frac{3}{4}\log[(x_1-x_2)(x_3-1)]+\phi(x_2).
\end{equation}
\subsection{Pure state limit}
\begin{figure}[H]
  \begin{center}
  \begin{tikzpicture}
  \draw[thick](-0.9,0)--(-0,0); 
  \draw[thick](1.8,0)--(2.7,0); 
  \draw[Nblue,thick](-5,0)--(-0.9,0);
  \draw[red,thick ](0,0)--(1.8,0);
  \draw[Nblue,thick](2.7,0)--(6.2,0);
  \draw(-0.9,0.3)--(-0.9,-0.3);
  \draw(0.0,0.3)--(0.0,-0.3);
  \draw(1.8,0.3)--(1.8,-0.3);
  \draw(2.7,0.3)--(2.7,-0.3);
  \draw(-5,0.3)--(-5,-0.3);
  \draw(6.2,0.3)--(6.2,-0.3);
  \node()at (-5.5,-0.5){$ \infty\leftarrow z_1 $};
  \node()at (-3.0,0.5){$B_1$};
  \node()at (-0.9,-0.5){$z_2$};
  \node()at (0.8,0.5){$A$};
  \node()at (3.8,0.5){$B_2$};
  \node()at (0.0,-0.5){$z_3$};
  \node()at (1.8,-0.5){$z_4$};
  \node()at (2.7,-0.5){$z_5$};
   \node()at (6.75,-0.5){$z_6\to\infty$};
  \end{tikzpicture}
  \end{center}
  \caption{Schematic of the configuration with single interval $A$ in between two disjoint intervals $B_1$ and $B_2$ in the limit of $B\to A^c$ (Pure state limit). }
  \label{fig6}
\end{figure}
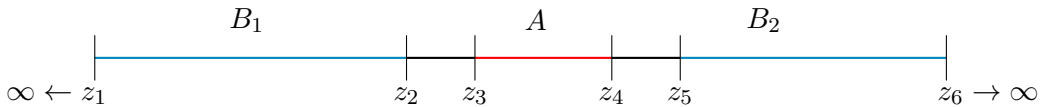
 It is important emphasize here that in order to fix the function $\phi(x_2)$ in eq.(\ref{fullf}) we require the third intermediate operator which leads to the third accessory parameter $c_3$. We observe that for this may be fixed only in the limit\footnote{Note that the points $z_1$ and $z_6$ are on either side of the interval $A$ and hence, the bipartite pure state limit  implies $z_1,z_6\to\infty$ as shown in fig.(\ref{fig6}). As the complex plane may be  mapped to a Riemann sphere on which infinity is a single point, this limit is equivalent to $z_1\to z_6$. This issue becomes extremely significant for the computation of the entanglement negativity for the bipartite finite temperature mixed state, where the bipartite limit is not equivalent to $z_1\to z_6$ as described in \cite{Calabrese:2014yza}. } $z_1\to z_6$ or $x_1,x_2,x_3\to 1$. The function $T(z)$ near $z=x_1=x_2=x_3=1$ in eq.(\ref{stress6}) may be approximated as follows
\begin{equation}\label{Tzc3}
T(z)=\frac{c_2 +c_3+c_4-c_2 x_1-c_3 x_2-c_4 x_3}{(z-1)^2}.
\end{equation}
The operators that fuse in the limit $z_1\to z_6$ are ${\cal T}_{n_e}(z_1)$ and $\overline{{\cal T}}_{n_e}(z_6)$ and therefore the vacuum block gives the dominant condition. Hence on utilizing the monodromy condition given by eq.(\ref{mono61}) with $h_p=0$ which corresponds  to the vacuum block, we obtain the accessory parameter $c_3$ as
\begin{equation}
 c_3=\frac{\partial f}{\partial x_2}=\frac{3 \left(2 x_1-x_2-1\right)}{4 \left(x_2-1\right). \left(x_2-x_1\right)}.\label{accpure}
\end{equation}
Upon solving the coupled partial differential equations given by eq.(\ref{acc6x1x2}), eq.(\ref{acc6x31}) and eq.(\ref{accpure}), we obtain the following form for the required function $f(x_1,x_2,x_3)$ in the pure state limit as
\begin{equation}\label{puref}
 f(x_1,x_2,x_3)=\frac{3}{4}\log\big[\frac{(x_1-x_2)(x_3-1)}{(x_2-1)^2}\big].
\end{equation}
Comparing eq.(\ref{accpure}) and  eq.(\ref{puref}), we obtain the function $\phi(x_2)$ in the limit of $x_2\to1$ as
\begin{equation}\label{x_2to1}
 \lim_{x_2\to1} \phi(x_2)=-\frac{3}{2}\log[(x_2-1)].
\end{equation}
From eq.(\ref{cr}), eq.(\ref{puref}), by taking $z_2-z_3=\epsilon$ and $z_4- z_5=\epsilon$ ( which corresponds to the limit of $B_1$ and $B_2$ becoming adjacent to $A$ ), we obtain the entanglement negativity as
\begin{equation}
{\cal E}=\frac{c}{2}\log[\frac{l_2}{\epsilon}]+O[\epsilon].
\end{equation}
The above expression matches exactly with the replica technique result given in eq.(\ref{Enpu}) and also that obtained by our conjecture given by eq.(\ref{Rhalf}), in the pure state limit.
\subsection{Adjacent intervals limit}
\begin{figure}[H]
	\begin{center}
		\begin{tikzpicture}
		\draw[thick](-4,0)--(-1.5,0); 
		\draw[thick](-1.5,0)--(0,0); 
		\draw[thick](1.8,0)--(2.7,0); 
		\draw[Nblue,thick](-1.5,0)--(-0.9,0);
		\draw[red,thick ](0,0)--(1.8,0);
		\draw[Nblue,thick](2.7,0)--(4.8,0);	
		\draw[thick](4.8,0)--(6.5,0);	
		\draw(-0.9,0.3)--(-0.9,-0.3);
		\draw(0.0,0.3)--(0.0,-0.3);
		\draw(1.8,0.3)--(1.8,-0.3);
		\draw(2.7,0.3)--(2.7,-0.3);
		\draw(-1.5,0.3)--(-1.5,-0.3);
		\draw(4.8,0.3)--(4.8,-0.3);
		\node()at (-1.5,-0.5){$ z_1 $};
		\node()at (-1.2,0.5){$B_1$};
		\node()at (-0.9,-0.5){$z_2$};
		\node()at (0.8,0.5){$A$};
		\node()at (3.8,0.5){$B_2$};
		\node()at (0.0,-0.5){$z_3$};
		\node()at (1.8,-0.5){$z_4$};
		\node()at (2.7,-0.5){$z_5$};
		\node()at (4.8,-0.5){$z_6$};
		\end{tikzpicture}
	\end{center}
	\caption{Schematic of the configuration with single interval $A$ in between two disjoint intervals $B_1$ and $B_2$, in the limit $B_1\to \emptyset$ }
	\label{fig7}
\end{figure}
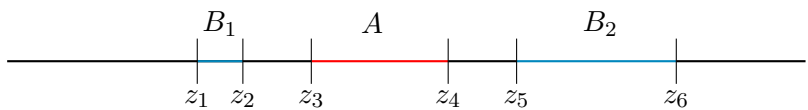
In the pure state limit $z_1\to z_6$ 
($x_2\to 1$) which was described above, we clearly demonstrated that the form of the  required function $\phi(x_2)$ in eq.(\ref{fullf}) is fixed . We will now utilize the large central charge limit of the entanglement negativity for the mixed state of the adjacent intervals  \cite{Kulaxizi:2014nma}, which was discussed in section.(\ref{secadj}), to further constrain the function $\phi(x_2)$. Note that the mixed state depicted in fig.(\ref{fig5}) reduces to the mixed state of the adjacent intervals depicted in fig.(\ref{fig7}) in the limit of the interval $B_1\to \emptyset$ or $z_1\to z_2$ ($x_1\to 0$ from eq.(\ref{cr}) and since  $x_1\to x_2$ this implies $x_2\to0$ as well ). Hence, in this limit the function $f(x_1,x_2,x_3) $ given by eq.(\ref{fullf}) has to reduce to the function given by eq.(\ref{fadj}).
This implies that
\begin{equation}\label{x2adj}
\lim_{x_1,x_2\to0}f(x_1,x_2,x_3)=\frac{3}{4}\log[\frac{(x_3-1)}{(x_2-1)}]\approx \frac{3}{4}\log[(x_3-1)]
\end{equation}
It may be easily checked from eq(\ref{cr}), that this form of the function leads to the following expression for the entanglement negativity in the adjacent interval limit ($z_1-z_2=\epsilon$, $z_2-z_3=\epsilon$ and $z_4-z_5=\epsilon$) 
\begin{equation}
{\cal E}=\frac{c}{4}\log\big[\frac{l_2l_3}{(l_2+l_3)\epsilon}\big]+O[\epsilon]
\end{equation}
Notice that as expected, the above expression matches exactly with the result obtained by \cite{Kulaxizi:2014nma} given in eq.(\ref{enlcadj}) and the replica technique result obtained by \cite{Calabrese:2012ew,Calabrese:2012nk} in eq.(\ref{enadj1}). From eq.(\ref{x2adj}) and eq.(\ref{fullf}), this in turn implies that the function $\phi(x_2)$ is fixed in the limit $x_2\to0$ as
\begin{equation}\label{x_2to0}
\lim_{x_2\to0}\phi(x_2)=-\frac{3}{4}\log[x_2].
\end{equation}
Hence, from eq.(\ref{x_2to1}) and eq.(\ref{x_2to0}), the  form of the function $\phi(x_2)$ is fixed in the limits $x_2\to1$ and $x_2\to0$. However, note that for the required mixed state configuration depicted in fig.(\ref{fig5}),  $x_1$ and $x_2$ can be anywhere between $0$ and $1$. The simplest form of the function $\phi(x_2)$ which reduces to eq.(\ref{x_2to1})  in the limit $x_2\to 1$ and eq.(\ref{x_2to0}) in the limit $x_2\to 0$ is as follows
\begin{equation}\label{phifu}
\phi(x_2)=-\frac{3}{4}\log[x_2(x_2-1)^2].
\end{equation}
Notice that the above choice of the function $\phi(x_2)$ obeys the constraints eq.(\ref{x_2to1}) and eq.(\ref{x_2to0}). We mention here that our choice of the function 
$\phi(x_2)$ remains non unique upto a function which conforms to the above mentioned constraints. However as we shall show below this choice of the function conforms to the 
bulk geodesic structure following from our holographic conjecture and obeys the required monodromy conditions. Seemingly the remnant non uniqueness in the choice of the function $\phi(x_2)$ is a limitation of the monodromy technique for this limit.

From eq.(\ref{phifu}) and eq.(\ref{fullf}) we get the form of the function $f(x_1,x_2,x_3)$ and the accessory parameters to be as follows
\begin{eqnarray}
f(\epsilon_p,\epsilon_i,x_1,x_2,x_3)&=&\frac{3}{4}\log\big[\frac{\left(x_1-x_2\right) \left(x_3-1\right)}{x_2 \left(x_2-1\right){}^2}\big],\label{fxfu}\\
 \frac{\partial f}{\partial x_1}&=& \frac{3}{4 (x_1 - x_2)},\label{fx1x2}\\
 \frac{\partial f}{\partial x_2}&=&\frac{6 x_2^2+x_1 \left(3-9 x_2\right)}{4 \left(x_1-x_2\right) \left(x_2-1\right) x_2},\label{fun}\\
 \frac{\partial f}{\partial x_3}&=&\frac{3}{4 \left(x_3-1\right)}.\label{fx31}
\end{eqnarray}
Utilizing the values of cross ratios given in eq.(\ref{cr})) in eq.(\ref{fxfu}),by taking $z_2-z_3=\epsilon$ and $z_4- z_5=\epsilon$ ( which corresponds to the limit of $B_1$ and $B_2$ becoming adjacent to $A$ ), we obtain the entanglement negativity from eq.(\ref{entG6}) and eq.(\ref{G6exp}) as
\begin{equation}
{\cal E}=\frac{c}{4}\log \Big[\frac{l_1~l_2^2 ~l_3}{(l_1+l_2)(l_2+l_3)\epsilon^2}\Big]+O[\epsilon].
\end{equation}
Once again upon choosing $\epsilon=a$ the above expression matches exactly  from
the holographic conjecture (CMS) for the entanglement negativity for this mixed state configuration obtained through the bulk pure $AdS_3$ geodesics given in eq.(\ref{Enrlc}) and  with the universal part of the entanglement negativity determined though the replica technique results given in eq.(\ref{Enrmix}). The above result therefore serves as a strong consistency check for the holographic conjecture proposed in \cite{Chaturvedi:2016rcn}.

Furthermore, we have also numerically verified through a mathematica code\footnote{We have modified the program for determining the trace of the monodromy matrix for the correlation function of twist fields corresponding to entanglement entropy given in\cite{Hartman:2013mia} by considering the appropriate intermediate operators ( $\overline{{\cal T}}_{n_e}^2$ and ${\cal T}_{n_e}^2$) for the required correlation function in eq.(\ref{sixG}).} by solving the differential equation given by eq.(\ref{ODE})  and then obtaining the trace of monodromies around the singularities of the stress tensor $T(z)$ in the limits $x_1\to x_2$ and $x_3\to 1$. Remarkably, with the  above mentioned accessory parameters, the monodromy conditions match exactly with the required conditions in eq.(\ref{mono6}). We have provided below a table describing our numerical computation of the trace of the required monodromy matrices for some representative values of $x_1,x_2$ and $x_3$ such that $x_1\to x_2$ and $x_3\to 1$
\begin{center}

		\begin{tabular}{ | m{0.2cm} | m{1cm}| m{1cm} |  m{1cm} | m{2cm} | m{2cm} | } 
			\hline
			& $x_1$ & $x_2$ & $x_3$	&$Tr(\textbf{M}_{x_1 x_2})$ & $Tr(\textbf{M}_{x_1 x_2})$\\ 
			\hline
		1&015&0.17&0.97& -2.006&-1.997\\
			\hline
		2 &0.24& 0.26&0.97&-2.007&-1.997 \\ 
			\hline
			3&0.32&0.35&0.96&-2.013&-1.992\\
			\hline
			4& 0.46 & 0.48 & 0.98&-2.007&-1.984\\ 
			\hline
			5& 0.51 & 0.54 & 0.99&-2.004&-1.999\\ 
			\hline
		\end{tabular}
\end{center}

\noindent Observe from the table above that the values of the trace of the required monodromy matrices match significantly with the expected monodromy conditions in eq.(\ref{mono6}) providing a strong consistency check for the validity of the holographic conjecture in \cite{Chaturvedi:2016rcn,Chaturvedi:2016rft}. Hence from the analytic and numerical determination of the large central charge behaviour of the six point function and the four point function corroborates the holographic entanglement negativity conjecture described in \cite{Chaturvedi:2016rcn,Chaturvedi:2016rft}.

\section{Summary and Conclusion}

To summarize, in this article we have utilized the monodromy technique to examine the
large central charge behavior of the entanglement negativity for a bipartite mixed state configuration in a $CFT_{1+1}$. The mixed state in question is described by a single interval enclosed between two intervals on either side in a holographic $CFT_{1+1}$. We would like to emphasize that this configuration is relevant to the recently proposed CMS conjecture for the holographic entanglement negativity of bipartite systems described by a $CFT_{1+1}$. To this end we have elucidated the large central charge limit of a  four point 
twist correlator characterizing the entanglement negativity of the relevant mixed state. Through this analysis we have 
explicitly demonstrated that the behaviour of the four point function both in $s$ and $t$ channels, is consistent with the CMS holographic negativity conjecture in the $AdS_3/CFT_2$ scenario.

We would like to point out here that the analysis described above for the four point twist correlator in different channels do not encompass the complete regime for the entanglement negativity of the relevant mixed state. To this end it was necessary to investigate the large central charge limit of a six point twist correlator in a specific colliding limit which captures the entire regime for the four point correlator mentioned above.  Utilizing the monodromy technique for the six point correlator we establish the form of the entanglement negativity for the required mixed state configuration, in the large central charge limit. We observe that universal part of the entanglement negativity is dominant
in the large central charge limit and conforms with the corresponding result obtained from our holographic conjecture. Furthermore we also solve the monodromy equation numerically and establish that that the expected monodromy conditions are precisely satisfied.
Naturally our analysis constitutes a strong consistency check for the  CMS holographic negativity conjecture in the context of the $AdS_3/CFT_2$ scenario. Our analysis also serves as a starting point for a bulk proof of the CMS holographic negativity conjecture. We hope to return to this fascinating issue in the near future.

\section {Acknowledgement } We would like to thank Ashoke Sen for crucial suggestions and Aninda Sinha for critical discussions and remarks. We acknowledge Tarek Anous, Arjun Bagchi, Shamik Banerjee, Pallab Basu, Ivano Lodato and R. Loganayagam for valuable criticisms and discussions during the {\it Strings Attached Workshop 2017} at IIT Kanpur, which inspired this work. We are grateful to {\it JoyGuru Pharmaceuticals} for continuing support and creative inspiration over the years.
\bibliographystyle{utphys} 
\bibliography{enlc} 
\end{document}